\documentclass[iop]{emulateapj}
\usepackage{amsmath}
\usepackage{times}
\usepackage{graphics, subfigure}
\usepackage{epsfig}
\usepackage{multirow}
\usepackage{ulem}
\usepackage{pdfpages}

\def\simlt{\mathrel{\hbox{\rlap{\hbox{\lower4pt\hbox{$\sim$}}}\hbox{$<$}}}}
\def\simgt{\mathrel{\hbox{\rlap{\hbox{\lower4pt\hbox{$\sim$}}}\hbox{$>$}}}}

\begin{document}

\title{Head-on collisions of white dwarfs in triple systems could explain type Ia supernovae}

\author{Doron Kushnir\altaffilmark{1,2}, Boaz Katz\altaffilmark{1}, Subo Dong\altaffilmark{1,3}, Eli Livne\altaffilmark{4},
and Rodrigo Fern\'andez\altaffilmark{1}} \altaffiltext{1}{Institute for Advanced Study, Einstein Drive, Princeton, NJ, 08540, USA} \altaffiltext{2}{Corresponding author, kushnir@ias.edu}\altaffiltext{3}{Current address: Kavli Institute for Astronomy and Astrophysics, Peking University, Yi He Yuan Road 5, Hai Dian District, Beijing 100871, China}\altaffiltext{4}{Racah Institute of Physics, Hebrew University, Jerusalem, Israel}

\begin{abstract}
Type Ia supernovae (SNe Ia), thermonuclear explosions of carbon-oxygen white dwarfs (CO-WDs), are currently the best cosmological ``standard candles'', but the triggering mechanism of the explosion is unknown. It was recently shown that the rate of head-on collisions of typical field CO-WDs in triple systems may be comparable to the SNe Ia rate. Here we provide evidence supporting a scenario in which the majority of SNe Ia are the result of such head-on collisions of CO-WDs. In this case, the nuclear detonation is due to a well understood shock ignition, devoid of commonly introduced free parameters such as the deflagration velocity or transition to detonation criteria. By using two-dimensional hydrodynamical simulations with a fully resolved ignition process, we show that zero-impact-parameter collisions of typical CO-WDs with masses $0.5-1\,M_{\odot}$ result in explosions that synthesize $^{56}$Ni masses in the range of $\sim0.1-1\,M_{\odot}$, spanning the wide distribution of yields observed for the majority of SNe Ia. All collision models yield the same late-time ($\simgt 60$ days since explosion) bolometric light curve when normalized by $^{56}$Ni masses (to better than $30\%$), in agreement with observations. The calculated widths of the $^{56}$Ni-mass-weighted-line-of-sight velocity distributions are correlated with the calculated $^{56}$Ni yield, agreeing with the observed correlation. The strong correlation, shown here for the first time, between $^{56}$Ni yield and total mass of the colliding CO-WDs (insensitive to their mass ratio), is suggestive as the source for the continuous distribution of observed SN Ia features, possibly including the Philips relation.
 \end{abstract}


\keywords{hydrodynamics ---  methods: numerical --- supernovae: individual (Ia)}

\section{Introduction}
\label{sec:Introduction}

There is strong evidence that Type Ia supernovae (SNe Ia) are thermonuclear explosions of carbon-oxygen white dwarfs \citep[CO-WDs;][]{Hoyle60}.
It is widely thought that the explosion is caused by accretion of matter onto the CO-WD, and as it approaches the unstable Chandrasekhar mass limit of $\approx1.4M_{\odot}$, a poorly understood mechanism causes the required explosion \citep[see][for a review]{hillebrandt2000typeIa}. In these models, the uncertainties in the burning waves allow the introduction of free parameters (such as the deflagration velocity or transition to detonation criteria), tuned to account for the observations. It was recently shown that the rate of head-on collisions of typical field WDs in triple systems may be as high as that of SNe Ia \citep[][see, however, \citet{Hamers2013}]{katz2012rate}, and it was suggested that some or all SNe Ia are due to such collisions. Here we provide evidence supporting a scenario in which the majority of SNe Ia are the result of such head-on collisions of CO-WDs. In this case the nuclear detonation is due to a well understood shock ignition, devoid of the commonly introduced free parameters, and unrelated to the Chandrasekhar limit.

Collisions of WDs were previously suggested to occur mainly in dense stellar systems
\citep[e.g., ][]{Hut1985, Sidgurdsson&Phinney93,Thompson11} \footnote{\citet{Thompson11} speculated about field triple systems that ``in some cases something akin to a `collision' or at least a very strong tidal interaction occurs.''}. While such collisions are believed to have rates which are orders of magnitude smaller than the rate of SNe Ia, they motivated three-dimensional (3D) hydrodynamic simulations of the thermonuclear explosion of colliding WDs \citep{Benz1989,Raskin2009oti,Rosswog2009cwd,Loren-Aguilar2010,Raskin2010pdd,Hawley2012zip}. While the amount of $^{56}$Ni \citep[the decay of which powers the observed light;][]{Colgate69} synthesized in most of these simulations was non-negligible, some results were contradictory, with inconsistent amounts of $^{56}$Ni and different ignition sites for the same initial conditions.

Section~\ref{sec:numerical} describes our simulations of zero-impact-parameter CO-WD collisions, which fully resolve the ignition process. In Section~\ref{sec:previous}, we compare our results to previous works, resolving previous discrepancies. Section~\ref{sec:observations} discusses several observational tests, which avoid radiation transfer uncertainties, and which demonstrate the consistency of our models with the majority of SNe Ia. Section~\ref{sec:future} presents avenues for future research.

\section{Numerical simulations}
\label{sec:numerical}

We consider the collisions of (equal and non-equal mass) CO-WDs with masses 0.5, 0.6, 0.7, 0.8, 0.9, 1.0$\,M_{\odot}$, covering the range of CO-WD masses. This problem is axisymmetric, allowing the use of two-dimensional (2D) numerical simulations with high resolution ($\sim$few km cell size), which is higher by at least an order of magnitude than those in previous Eulerian approaches \citep[limited to $>100$ km cell sizes;][]{Hawley2012zip} and smooth particle hydrodynamic (SPH) models \citep[][in which the resolution is estimated from the ``cell volume'', corresponds to particle mass divided by the density]{Raskin2010pdd}. We employ two different hydrodynamic codes: FLASH4.0 \citep[][Eulerian, adaptive mesh refinement]{dubey2009flash} and VULCAN2D \citep[][arbitrary Eulerian-Lagrangian, ALE]{Livne1993IMT}. FLASH4.0 solves the equations of reactive hydrodynamics by combining the dimensionally-split Piecewise Parabolic Method \citep{Colella84,Fryxell2000} with a 19 isotope alpha-chain reaction network \citep{Timmes1999}. The system of equations is closed with the Helmholtz equation of state \citep{Timmes2000} and a multipole gravity solver. VULCAN2D solves the reactive hydrodynamic equations on an ALE grids using a Lagrangian step followed by a remapping step on a new grid with a 13 isotope alpha-chain reaction network. We exploit the flexibility of the grid to refine it near the collision interface, so that at the onset of a detonation a maximal, pre-determined resolution is reached there. The equation of state, gravity solver and boundary conditions are similar to those used in FLASH4.0. Unless otherwise stated, we present the FLASH4.0 results.

Initially the CO-WDs are at contact with free fall velocities. The structure of each CO-WD is obtained from an isothermal stellar model\footnote{http://cococubed.asu.edu/code\_pages/adiabatic\_white\_dwarf.shtml} at $T=10^7$~K and with a uniform composition of $50\%$ carbon and $50\%$ oxygen by mass.

\begin{turnpage}
\begin{figure}
\includegraphics[width=22.0cm]{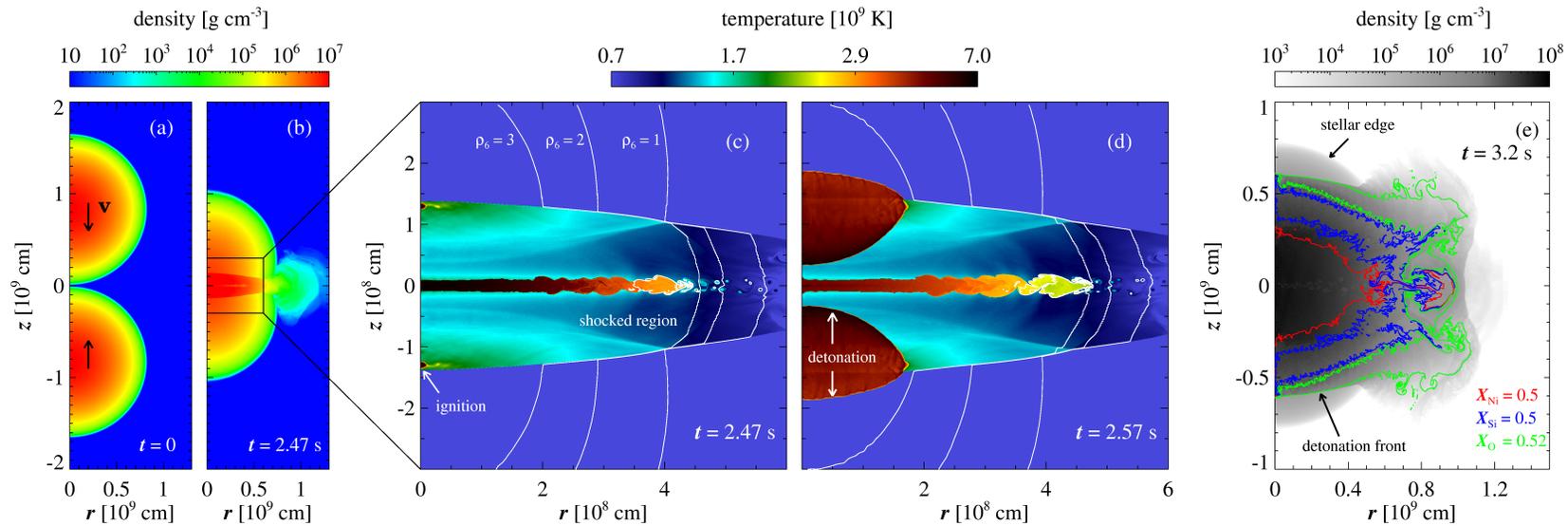}
\caption{Snapshots from a FLASH4.0 simulation of a $0.64-0.64\,M_{\odot}$ collision ($\simeq4\,\textrm{km}$ resolution). Panel (a): a logarithmic density map of the initial conditions. Black arrows indicate the direction of the velocity (assumed uniform) of each CO-WD. Following contact, shock waves propagate from the contact surface toward each star's center. The shocks accelerate and detonation ignition occurs once the post-shock induction time is shorter than the timescale for significant increase in burning rate \citep{zeldovich}. Panel (b): density map at the time of ignition ($t=2.47\,\textrm{s}$). Panel (c): temperature map with density contours, showing the ignition sites in each star. Panel (d): same as panel (c), but showing evolution of the twin detonation waves at $t=2.57\,\textrm{s}$. Panel (e): density map with isotope contours at $t=3.2\,\textrm{s}$, showing stratified ejecta structure caused by detonations.
\label{fig:ignition 2D}}
\end{figure}
\end{turnpage}

\begin{figure}
        \subfigure[]{
             \includegraphics[width=0.5\textwidth]{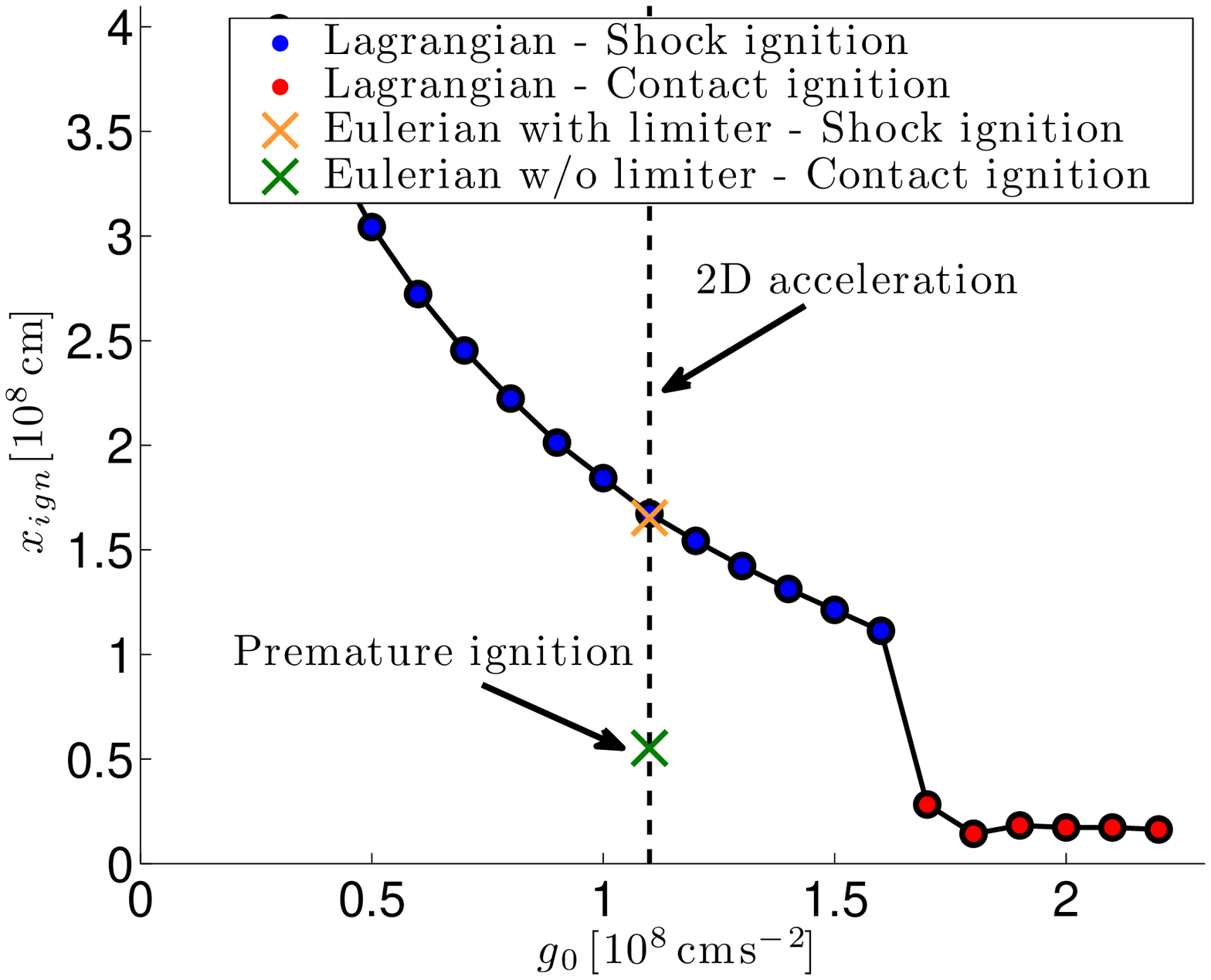}
        }
        \subfigure[]{
             \includegraphics[width=0.5\textwidth]{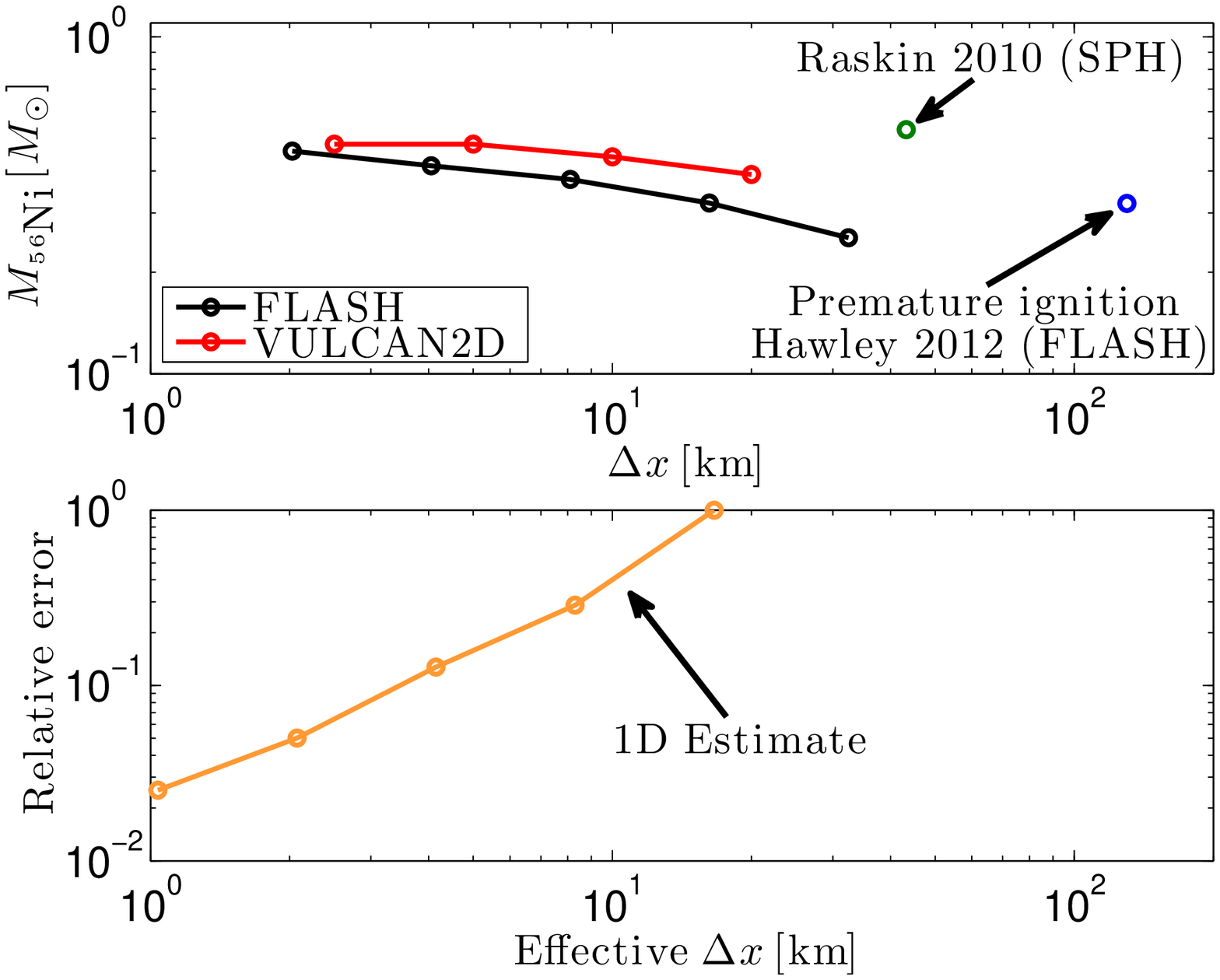}
        }
      \caption{Panel (a): the ignition process shown in Figure~\ref{fig:ignition 2D} is confirmed using a planar 1D model evolved with Lagrangian and Eulerian schemes. The converged location of the ignition, $x_{\rm ign}$, obtained with a high-resolution Lagrangian scheme, is shown as a function of the adjustable acceleration, $g_{0}$, for the $0.64-0.64\,M_{\odot}$ collision: blue (shock region ignition), red (contact region ignition). The acceleration $g_0\approx 1.1\times 10^8\,\textrm{cm}\,\textrm{s}^{-2}$, shown as a dashed line, approximately reproduces the 2D velocity profiles in Figure~\ref{fig:ignition 2D}. The burning in the Eulerian code with resolution $\Delta x\sim \textrm{few}\,\rm km$ is unstable (see text), and leads to a premature detonation at the contact region (green cross) unless the burning limiter is included, where the correct ignition location is reproduced (orange cross). Upper panel (b): the convergence of the $^{56}$Ni mass as a function of resolution for $0.64-0.64\,M_{\odot}$ collision model. Our Eulerian FLASH (blacks) and the ALE VULCAN2D (red) converge to the same value. The SPH calculation of \citet{Raskin2010pdd} for this case yields a similar result (green point). In the previous Eulerian calculation of \citet{Hawley2012zip}, premature ignition at the contact surface occurs (blue point), due to burning which is faster than the cell sound crossing time. Lower panel (b): the decreasing error in the $^{56}$Ni yield as a function of resolution in the 1D Eulerian model. Based on the 1D and the 2D runs, we estimate that the $^{56}$Ni yield in our high-resolution FLASH4.0 runs ($3-5$ km cell size) is converged to about $10\%$.
     }
   \label{fig:1D and Conv}
\end{figure}

Figure~\ref{fig:ignition 2D} shows an example of the ignition process and the ejecta formation for a $0.64-0.64\,M_{\odot}$ collision. For this case we verified that placing the CO-WDs with a separation of 4 stellar radii between their centers has a minor effect (a few percent) on our results. We used VULCAN2D to evolve several cases, and found the same $^{56}$Ni mass and the same location of the detonation ignition (see, e.g., the right panel of Figure~\ref{fig:1D and Conv}). In both codes, collisions involving CO-WDs with mass $\simlt0.7\,M_{\odot}$ ignite at the shock region (as in Figure~\ref{fig:ignition 2D}), while higher mass CO-WDs ignite at the contact region.

The shocked region at the vicinity of the symmetry axis has approximate planar symmetry (see, e.g., panel (c) of Figure~\ref{fig:ignition 2D}). This allowed us to develop a one-dimensional (1D) planar model (evolved with Lagrangian and Eulerian schemes) to ensure that 2D simulations correctly resolve the ignition process (panel (a) of Figure~\ref{fig:1D and Conv}). The initial CO-WD density of the 1D model equals the density on the axis of symmetry in the 2D model, and the initial velocity is the free fall velocity. The gravitational field is mimicked by an adjustable acceleration, which is constant in time and space. For example, the acceleration $g_0\approx 1.1\times 10^8\,\textrm{cm}\,\textrm{s}^{-2}$, approximately reproduces the 2D velocity profiles of the $0.64-0.64\,M_{\odot}$ collision in Figure~\ref{fig:ignition 2D}, and results in a similar location of the ignition ($\sim1.5\times 10^8\,\textrm{cm}$ from the contact surface). In particular, this suggests that the curvature of the shock has little effect on the ignition process. By artificially changing the acceleration we verify the robustness of the correct ignition location. Only for a substantially higher value of the acceleration $g_0\approx1.6\times 10^8\,\textrm{cm}\,\textrm{s}^{-2}$ does the detonation occur at the contact region.

Numerically unstable burning occurs if the energy in a cell is significantly increased in a time shorter than the sound crossing time, $t_{s}\equiv\Delta x/c_{s}$, where $\Delta x$ is the length scale of the cell and $c_{s}$ is the sound speed. This is a severe problem for Eulerian calculations with cell sizes larger than $\sim\textrm{km}$. Stability is achieved by limiting the energy injection rate from burning to $f\varepsilon/t_{s}$, where $\varepsilon$ is the internal energy of the cell and \textbf{$f\ll1$}. This is implemented in all of our simulations by appropriate renormalization of all burning rates within a cell if the limit is exceeded. As the numerical resolution increases, the renormalization becomes less severe, guaranteeing correct convergence while avoiding premature ignition. The limiter does not modify the ignition of a detonation wave process, which is set at lower temperatures, where the stability criterion is automatically satisfied. By comparing the 1D Eulerian model with resolution $\Delta x\sim \textrm{few}\,\rm km$ to a converged 1D Lagrangian model, we verify that the limiter doesn't affect the ignition and burning for $f\lesssim0.4$. Therefore, we adopted $f=0.1$ in our simulations. In several cases we find that if the limiter is not used, a premature numerical ignition may occur (see, e.g., left panel of Figure~\ref{fig:1D and Conv}).

The amount of $^{56}$Ni synthesized in the various collisions is shown in Figure~\ref{fig:NiMass} (and is given in Table~1), which is the main result of this Letter. We find that CO-WD collisions lead to the synthesis of $\sim 0.1 -1 M_{\odot}$ of $^{56}$Ni, covering the range of yields observed in the vast majority of SNe Ia (see, e.g., Figure~\ref{fig:nebular_vmod_ni56}). Furthermore, there is a tight correlation between the $^{56}$Ni yields and the total mass of the colliding CO-WD masses, which is insensitive to their mass ratio.

\section{Comparison to previous calculations}
\label{sec:previous}

For equal mass collisions, similar $^{56}$Ni yields were obtained by most of the SPH simulations of \citet{Rosswog2009cwd} and of \citet{Raskin2010pdd} (except for the $0.5-0.5\,M_{\odot}$ collision, for which both groups obtained a negligible amount of $^{56}$Ni), and significantly lower yields (by factor of $\sim2$) were obtained by FLASH calculations of the same groups \citep{Rosswog2009cwd,Hawley2012zip}. However, most non-equal mass SPH results contradict ours. Four such simulations exist, and three of them (the $0.9-0.6\,M_{\odot}$ collision of \citet{Rosswog2009cwd} and the $0.81-0.64\,M_{\odot}$ and $1.06-0.64\,M_{\odot}$ collisions of \citet{Raskin2010pdd}) have lower $^{56}$Ni yields than ours by factors of a few up to orders of magnitude. As a result, the tight correlation between the $^{56}$Ni yields and the total mass of the colliding CO-WD masses was not seen in any of the previous works.

We reproduce the low $^{56}$Ni yields of the discrepant simulations by running simulations with similarly low resolutions. We are able to identify the main cause for the discrepancy in most cases. In a $0.64-0.64\,M_{\odot}$ simulation with a similar setup as in \citet{Hawley2012zip}, a premature ignition occurs at the contact region that prevents the correct shock ignition at later times. This occurs mainly due to the fact that \citet{Hawley2012zip} did not implement the required burning limiter to avoid numerical unstable burning (see Section~\ref{sec:numerical}). We also performed $0.5-0.5\,M_{\odot}$, $0.81-0.64\,M_{\odot}$, and $0.9-0.6\,M_{\odot}$ simulations with similar resolutions to the cell volumes of the SPH simulations. We find that the propagation of the detonation wave is significantly altered due to the low resolution. For example, in our high-resolution $0.81-0.64\,M_{\odot}$ simulation, following ignition in the $0.64\,M_{\odot}$ star, the detonation wave propagates inside the $0.64\,M_{\odot}$ star in the $r$-direction up to a height of $\sim3000\,\textrm{km}$ from the symmetry axis, and then crosses into the $0.81\,M_{\odot}$ star. However, the detonation crossing, which is responsible for the large amount of $^{56}$Ni ($\sim0.4\,M_{\odot}$), is not resolved by \citet{Raskin2010pdd}, in which the resolution is lower than our high-resolution runs by a factor of $\sim100$ in the crossing location (density of $\textrm{few}\times10^{6}\,\textrm{g}\,\textrm{cm}^{-3}$). As a consequence, the low-resolution runs obtain small amounts of $^{56}$Ni. Given that similar dynamics are observed in our $1.0-0.6\,M_{\odot}$ collision, we suspect that the low $^{56}$Ni yield obtained for the $1.06-0.64\,M_{\odot}$ collision of \citet{Raskin2010pdd} is due to the same reason.

\section{Comparison to observations}
\label{sec:observations}

\begin{figure}
\includegraphics[width=16.0cm]{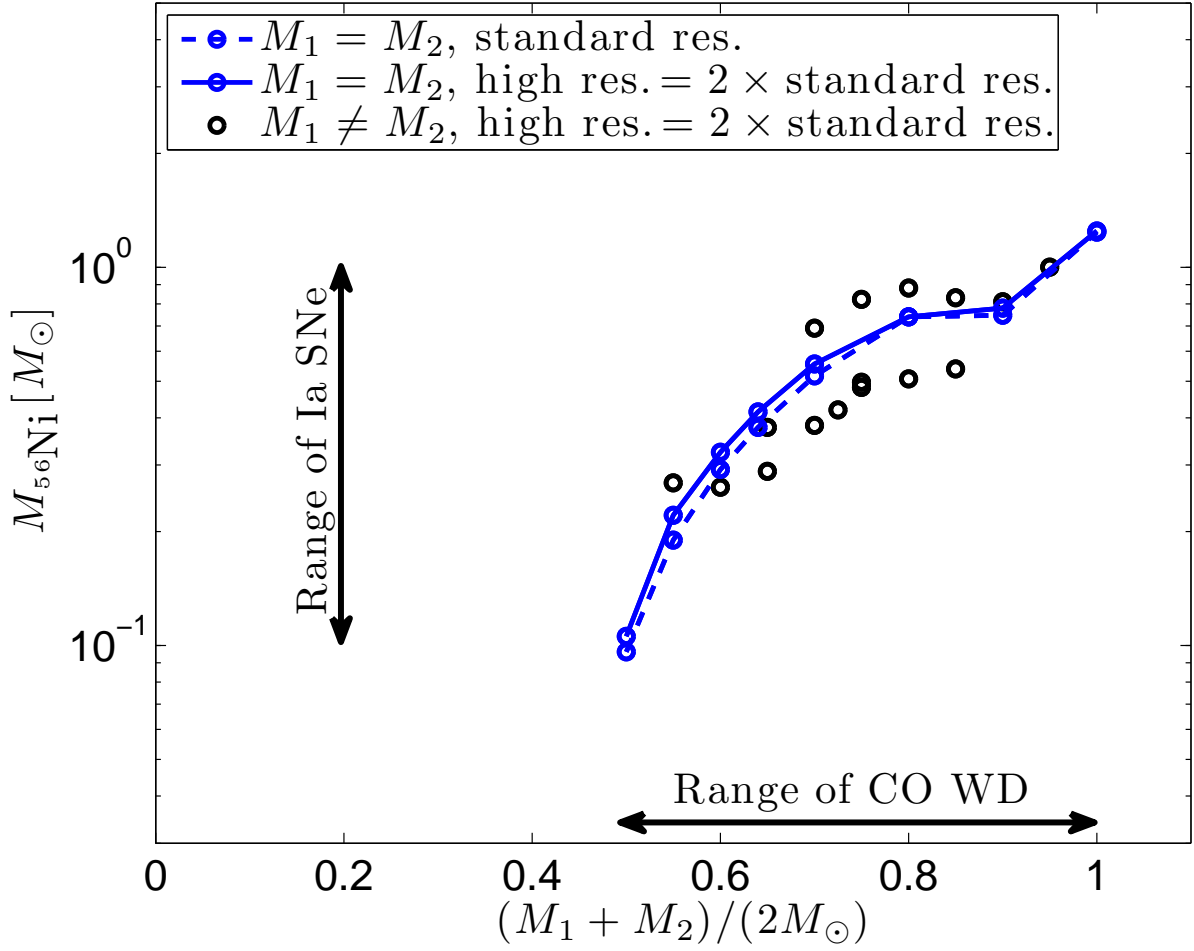}
\caption{ Synthesized $^{56}$Ni mass as a function of the colliding CO-WDs masses. The amount of $^{56}$Ni synthesized in the simulated collisions is shown as a function of the mean mass of the colliding CO-WDs at two resolutions (solid and dashed) and for equal (blue) and non-equal mass (black) collisions. The high-resolution data in solar units is ($M_{1}-M_{2}-M_{^{56}\textrm{Ni}}$): 0.5-0.5-0.11; 0.55-0.55-0.22; 0.6-0.5-0.27; 0.6-0.6-0.32; 0.64-0.64-0.41; 0.7-0.5-0.26; 0.7-0.6-0.38; 0.7-0.7-0.56; 0.8-0.5-0.29; 0.8-0.6-0.38; 0.8-0.7-0.48; 0.8-0.8-0.74; 0.81-0.64-0.42; 0.9-0.5-0.69; 0.9-0.6-0.50; 0.9-0.7-0.51; 0.9-0.8-0.54; 0.9-0.9-0.78; 1.0-0.5-0.82; 1.0-0.6-0.88; 1.0-0.7-0.83; 1.0-0.8-0.81; 1.0-0.9-1.00; 1.0-1.0-1.25.
\label{fig:NiMass}}
\end{figure}

\begin{figure}
\includegraphics[width=12.5cm]{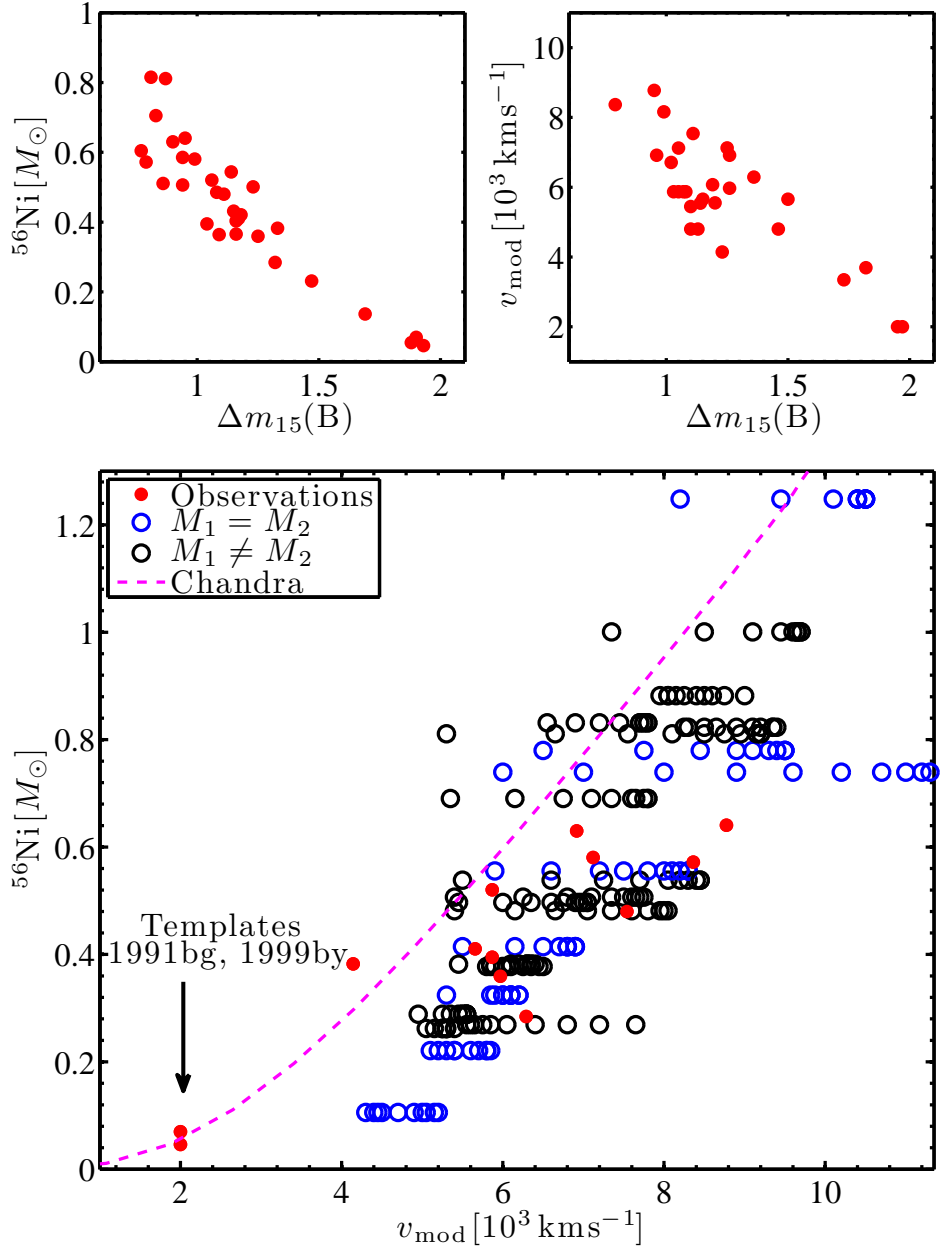}
\caption{Nebular-phase velocities. Bottom panel: comparison between the distribution of widths of the ($^{56}$Ni mass-weighted) LOS velocities ($v_{\textrm{mod}}$, defined in the text) for the collision models (empty circles) and those inferred from nebular-phase observations (red). For purposes of illustration, the simple Chandrasekhar-model prediction is shown in magenta. Top panels: larger samples are used to show the continuous correlation of both $^{56}$Ni mass and $v_{\textrm{mod}}$ with $\Delta m_{15}(B)$ (the decline in $B$-band magnitude during the first $15$ days after peak), used to establish the Philips relation \citep{phillips}. The typical errors of the observed $^{56}$Ni mass and $v_{\textrm{mod}}$ are $\sim30\%$.
\label{fig:nebular_vmod_ni56}}
\end{figure}

The amount of $^{56}$Ni obtained in our simulations (Figure~\ref{fig:NiMass}), implies that collisions of typical CO-WDs in the mass range $0.5-1\,M_{\odot}$ produce $^{56}$Ni masses in the range $0.1-1\,M_{\odot}$ in agreement with the inferred range of SNe Ia (see, e.g., Figure~\ref{fig:nebular_vmod_ni56}). The tight correlation between the $^{56}$Ni yields and the total mass of the colliding CO-WD masses, which is insensitive to their mass ratio, may explain the fact that many of the SNe Ia properties are tightly correlated.

One constraint for explosion models is the observed substantial amount of high-velocity intermediate mass elements in the outer layers of the ejecta. Such elements are naturally produced in the collisions \citep[e.g.,][]{Rosswog2009cwd,Raskin2010pdd}, as can be seen in Figure~\ref{fig:ignition 2D}.

Our models pass two additional quantitative, nontrivial, and robust observational tests, which are independent of the complicated optical radiation transfer.

\begin{figure}
\includegraphics[width=14.0cm]{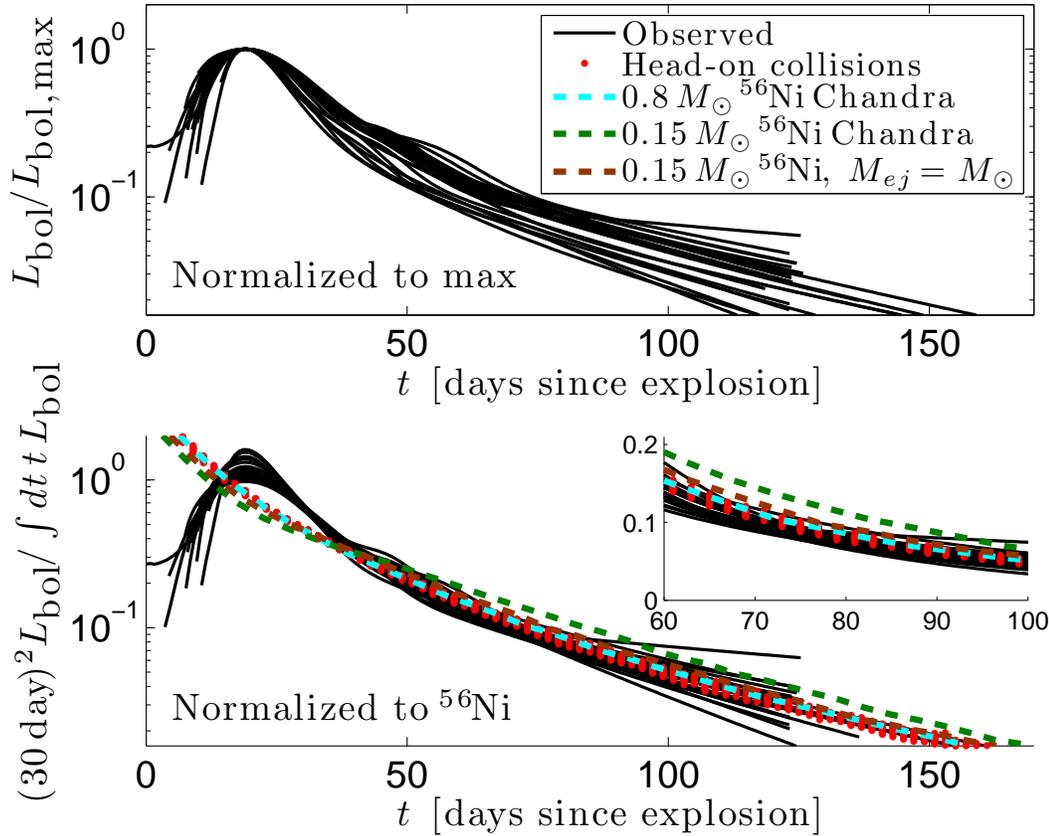}
\caption{Normalized bolometric light curves. The observed sample (black) is normalized by the peak luminosity (upper panel) and by the (time-weighted) integrated luminosity, $\int dt t L(t)$ \citep[proportional to the $^{56}$Ni yield;][lower panel]{katz2013Ni56}. The integral is performed to 80 days since explosion (assumed to be 19 days prior to maximum). The (volume-integrated) energy deposition from $\gamma$-rays and positrons $Q(t)$, normalized by $\int dt t Q(t)$, for all collision models are shown in red. At late times, $\simgt 40\,\rm days$, the diffusion of optical radiation is short and the energy deposition equals the emitted bolometric luminosity (integrated over viewing angles). The late-time luminosity of the models is in excellent agreement with the observed light curves (the inset contains a close up view on $60<t<100\,\textrm{days}$). This agreement is achieved without any fitting of the models to the observations. The late-time luminosities calculated from simple, spherical models (Chandrasekhar-model with $^{56}$Ni yields of $0.8 M_{\odot}$ (cyan) and $0.15 M_{\odot}$ (green), and $1.0\,M_{\odot}$ ejecta with $^{56}$Ni yield of $0.15 M_{\odot}$ (brown)) are shown for comparison (see text for discussion). Note that a much larger scatter is obtained when the light curves are normalized to the peak luminosity, indicating that it is not as an accurate estimator for the $^{56}$Ni yield as commonly believed \citep[``Arnett's rule'';][]{Arnett1979,Arnett1982}. \label{fig:Late_Bolometric}}
\end{figure}

The main channel by which the radioactive decay chain $^{56}\textrm{Ni}\rightarrow^{56}\textrm{Co}\rightarrow^{56}\textrm{Fe}$ converts its energy to observed light, is through the energy deposition of $\gamma$-ray photons and positrons into the ejecta. The heated ejecta reprocesses this energy into observed optical light. The late-time ($\simgt60$ days since explosion) bolometric luminosity equals the instantaneous $\gamma$-ray energy deposition rate, which is calculated by a Monte Carlo code for the Compton-scattering-dominated transport of $\gamma$-rays (with a small contribution from positrons which are assumed to deposit their kinetic energy locally). The ejecta is taken from the calculations at a sufficiently late time where the expansion is homologous. We find that the deposited fraction of $\gamma$-rays as a function of time is the same in all of our models (within $30\%$, where our results converge to better than the numerical statistical noise of $\sim3\%$\footnote{Note that the deposited fraction is converged to a better level than the $^{56}$Ni mass.}) and in excellent agreement with the sample of $24$ low extinction (Galactic+host $E(B-V)<0.3$) bolometric light curves of \citet{stritzinger2005thesis}, when appropriately normalized by the (time-weighted) integrated luminosity (see Figure~\ref{fig:Late_Bolometric}). Given that the ejecta is not spherically symmetric, the emission may depend on viewing angle, possibly widening the scatter of expected late-time luminosities. While this possible correction is beyond the scope of this Letter, the angle-average value should be accurately represented by our results.

To illustrate that this agreement is nontrivial, we calculated the light curves of simple Chandrasekhar toy models \citep[see, e.g.,][]{Jeffery92,Nugent95,Woosley2007}, with an exponential density distribution as function of velocity ($\rho\propto e^{-v/v_{e}}$, where $v_{e}$ is set by the total release of energy in the explosion) and varying $^{56}$Ni masses ($0.15-0.8\,M_{\odot}$). While the late-time light curve of the $0.8\,M_{\odot}$ $^{56}$Ni Chandrasekhar-model lies at the high end of observed light curves, the $0.15\,M_{\odot}$ $^{56}$Ni Chandrasekhar-model has a much higher normalized light curve and is inconsistent with observations (see Figure~\ref{fig:Late_Bolometric}). The strong emission at late time is due to the efficient deposition of $\gamma$-rays in the slowly moving, massive $1.25\,M_{\odot}$ ejecta surrounding the $^{56}$Ni. To confirm that the main problem with the Chandrasekhar toy model is the high ejecta mass, we apply a ``collision-like'' spherical model. This model has the same exponential density distribution, but with a smaller ejecta mass ($1.0\,M_{\odot}$), roughly corresponding to the relevant head-on collision for the $0.15\,M_{\odot}$ $^{56}$Ni yield (the kinetic energy is set by the same collision). The ``collision-like'' model has a lower normalized light curve, and shows a much better agreement with the observations (Figure~\ref{fig:Late_Bolometric}). We note that Chandrasekhar-models in which the $^{56}$Ni mass is located far from the center may alleviate this particular problem. More details will be provided in a subsequent paper \citep{BolPaper}.

The distribution of widths of the ($^{56}$Ni-mass weighted) line-of-sight (LOS) velocities for the collision models is compared to those inferred from nebular-phase observations of SNe Ia in Figure~\ref{fig:nebular_vmod_ni56}. For each viewing angle, the model widths ($v_{\textrm{mod}}$) are obtained by fitting the distribution of $^{56}$Ni mass per unit LOS velocity in that angle using a quadratic function, $dM_{^{56}\textrm{Ni}}/dv_\textrm{LOS}\propto1-v_{\textrm{LOS}}^{2}/v_{\textrm{mod}}^{2}$ \citep[e.g.,][]{Mazzali98}. The observational widths are obtained by fitting the observed late-time ($>$150 days) spectra\footnote{Obtained from the Berkeley Supernova Ia Program  \citep[BSNIP;][]{Silverman2013}, the Center for Astrophysics Supernova Program \citep{Blondin2012}, and the compilation from various sources by the online Supernova Spectrum Archive (SUSPECT   http://nhn.nhn.ou.edu/$\sim$suspect/).} in the range $4800-5700\,\textrm{\AA}$ to the convolution of the low-width spectra templates of 1991bg and 1999by with the same quadratic velocity distribution. We note that while several properties of 1991bg and 1999by are uncommon, their nebular spectra in the above range are very similar to other SNe \citep[e.g.,][except for the line widths]{Mazzali98} and should suffice as templates for measuring the line widths of other SNe \citep[see][for more details]{NebularPaper}. The amount of $^{56}$Ni is obtained by fitting the bolometric light curves at $t\sim60\,\textrm{days}$ to the universal injection function presented in Figure~\ref{fig:Late_Bolometric}, which is well described by $L_{\textrm{deposit}}=(1+(t_{d}/40)^{3})^{-2/3}L_{\textrm{decay}}$, where $t_{d}$ is the time since explosion in days and $L_{\textrm{decay}}$ is the energy injection in $\gamma$-rays by $^{56}$Ni and $^{56}$Co. There is a clear correlation between the observed $^{56}$Ni masses and the nebular-phase velocity widths. Both the correlation and the scatter of this Mazzali relation \citep{Mazzali98} are well reproduced by the collision model (10 viewing angles for each calculation, equally spaced in $\cos(i)$; these widths are converged to a level of $\sim20\%$). As can be seen in Figure~\ref{fig:nebular_vmod_ni56}, the simple Chandrasekhar-model predicts a similar correlation, somewhat in offset. The amount of SNe Ia with nebular spectra and well described bolometric light curves is limited. In the top panels larger samples (given in Table~2) are used to show the continuous correlation of observational features of SNe Ia with $^{56}$Ni yields. The strong correlation between $^{56}$Ni yield and the total mass of the colliding CO-WDs, Figure~\ref{fig:NiMass}, is suggestive as the source for these correlations, possibly including the Philips relation \citep{phillips}.

\section{Directions for future research}
\label{sec:future}

A future robust and detailed test is the time dependent MeV scale $\gamma$-ray spectrum, which can be computed in a straightforward manner. Other tests, closely related to the nebular velocity widths, include the distribution of non-zero nebular line shifts, which are expected in the collision of non-equal mass CO-WDs, as well as the unique line shapes resulting from the nontrivial velocity distribution in many collision scenarios. Finally, the early-time spectra and light curves can be compared to observations using suitable radiation transfer models \citep[see encouraging attempts in][]{Rosswog2009cwd}.

We stress that our calculations were restricted to the zero-impact-parameter. Only if the effect of the non-zero-impact-parameter is small for a large range of impact parameters, do our calculations represent a significant fraction of collisions \citep[the distribution of impact parameters is uniform for collisions in triple systems;][e.g., $50\%$ of collisions have impact parameters $<0.5(R_{1}+R_{2})$, where $R_{1,2}$ are the radii of the CO-WDs]{katz2012rate}. \citet{Raskin2010pdd} showed that ignition of a detonation can be triggered for several non-zero-impact-parameter collisions in SPH simulations. However, their simulations may suffer from numerical problems which are similar to the ones we identified for their zero-impact-parameter runs in Section~\ref{sec:previous}, making these results  quantitatively uncertain. Non-zero-impact-parameter collisions should be studied using 3D hydrodynamical codes of similar accuracy to the 2D calculation presented here. High-resolution 3D calculations are also desirable for constraining 3D effects for the zero-impact-parameter collisions.

\acknowledgments We thank A. Gould, R. Kirshner, J. Prieto, M. Zaldarriaga, E. Waxman, and F. Dyson for discussions. D.K., S.D., and R.F. are supported by NSF grant AST-0807444. B.K. is a John N.\ Bahcall Fellow, and supported by NASA through the Einstein Postdoctoral Fellowship awarded by Chandra X-ray Center, which is operated by the Smithsonian Astrophysical Observatory for NASA under contract NAS8-03060. S.D. was supported through a Ralph E. and Doris M. Hansmann Membership at the IAS. FLASH was in part developed by the DOE NNSA-ASC OASCR Flash Center at the University of Chicago. Computations were performed at PICSciE and IAS clusters.


\bibliographystyle{apj}

\newpage

\begin{table}
\begin{center}
\bigskip
\begin{tabular}{cccccc}
\\
\hline
\hline

$M_{1}\,[M_{\odot}]$ & $M_{2}\,[M_{\odot}]$ & $\Delta x\,[\textrm{km}]$ & $\textrm{Ignition type}$ & $^{56}\textrm{Ni}\;\textrm{mass}\,[M_{\odot}]$ & $E_{\textrm{tot}}\,[10^{51}\,\textrm{erg}]$ \\
\hline
 0.5  & 0.5  & 4.8 & s     & 0.11 & 0.99 \\
                0.55 & 0.55 & 4.5 & s     & 0.22 & 1.16 \\
                0.6  & 0.5  & 4.8 & s2,s1 & 0.27 & 1.17 \\
                0.6  & 0.6  & 4.2 & s     & 0.32 & 1.32 \\
                0.64 & 0.64 & 4.1 & s     & 0.41 & 1.45 \\
                0.7  & 0.5  & 4.8 & c2,s1 & 0.26 & 1.18 \\
                0.7  & 0.6  & 4.2 & c2,s1 & 0.38 & 1.42 \\
                0.7  & 0.7  & 3.8 & s     & 0.56 & 1.64 \\
                0.8  & 0.5  & 4.8 & c2,s1 & 0.29 & 1.23 \\
                0.8  & 0.6  & 4.2 & c2,s1 & 0.38 & 1.45 \\
                0.8  & 0.7  & 3.8 & c2,s1 & 0.48 & 1.65 \\
                0.8  & 0.8  & 3.4 & c     & 0.74 & 1.91 \\
                0.81 & 0.64 & 4.1 & c2,s1 & 0.42 & 1.54 \\
                0.9  & 0.5  & 4.8 & c2,x1 & 0.69 & 1.54 \\
                0.9  & 0.6  & 4.2 & c2,x1 & 0.50 & 1.58 \\
                0.9  & 0.7  & 3.8 & c2,x1 & 0.51 & 1.69 \\
                0.9  & 0.8  & 3.4 & c2,s1 & 0.54 & 1.89 \\
                0.9  & 0.9  & 3.0 & c     & 0.78 & 2.11 \\
                1.0  & 0.5  & 4.8 & c2,s1 & 0.82 & 1.67 \\
                1.0  & 0.6  & 4.2 & c2,s1 & 0.88 & 1.84 \\
                1.0  & 0.7  & 3.8 & c2,s1 & 0.83 & 1.92 \\
                1.0  & 0.8  & 3.4 & c2,x1 & 0.81 & 2.05 \\
                1.0  & 0.9  & 3.0 & c2,s1 & 1.00 & 2.27 \\
                1.0  & 1.0  & 2.7 & c     & 1.25 & 2.45 \\
\hline
\end{tabular}
\end{center}
\label{tbl:summary}

\noindent{}{Table 1.-- ~Summary of the high-resolution numerical simulations performed with FLASH4.0. Columns 1,2: The masses of the CO-WDs ($M_{1}$ is the more massive CO-WD); Column 3: the size of cells in the highest refinement level; Column 4: The type of ignition obtained in the simulation: for equal mass collisions the ignition is either at the shock region (s) or at the contact region (c). For non-equal mass collisions, the less massive star always ignites first, either near the shock region (s2) or near the contact region (c2), followed either by an ignition of the more massive star at the shock region (s1) or no second ignition (x1); Column 5: $^{56}$Ni mass synthesized in the explosion; Column 6: The total energy of the ejecta.}

\end{table}

\begin{table}
\begin{center}
\bigskip
\begin{tabular}{cccccccc}
\hline
\hline
$\textrm{Name}$ & ${^{56}\textrm{Ni}}$ & $\Delta{m_{15}(B)}$ & $v_{\textrm{mod}}$
& $\textrm{Name}$ & ${^{56}\textrm{Ni}}$ & $\Delta{m_{15}(B)}$ & $v_{\textrm{mod}}$\\ & $[M_\odot]$ & & $[10^3\,\textrm{km/s}]$
&  & $[M_\odot]$ & & $[10^3\,\textrm{km/s}]$\\\hline

1972E & -- & -- & 7.3 & 1981B & -- & 1.1 & 4.8\\
1986G & -- & 1.7 & 3.3 & 1990N & -- & 1.0 & 5.9\\
1991M & -- & 1.5 & 5.7 & 1991T & 0.64 & 0.9 & 8.8\\
1991bg & 0.05 & 1.9 & 2.0 & 1992A & 0.23 & 1.5 & --\\
1993H & 0.14 & 1.7 & -- & 1993Z & -- & -- & 4.6\\
1994D & 0.28 & 1.4 & 6.3 & 1994ae & 0.63 & 1.0 & 6.9\\
1995D & 0.58 & 1.1 & 7.1 & 1995E & 0.40 & 1.2 & --\\
1995ac & 0.60 & 0.8 & -- & 1995al & 0.71 & 0.9 & --\\
1995bd & 0.51 & 0.9 & -- & 1996X & 0.36 & 1.3 & 6.0\\
1996bo & 0.37 & 1.2 & -- & 1997bp & 0.49 & 1.1 & --\\
1997bq & 0.54 & 1.1 & -- & 1998aq & 0.48 & 1.1 & 7.5\\
1998bu & 0.39 & 1.1 & 5.9 & 1998de & 0.05 & 1.9 & --\\
1999aa & 0.57 & 0.8 & 8.4 & 1999ac & 0.50 & 1.2 & --\\
1999aw & 0.82 & 0.8 & -- & 1999by & 0.07 & 2.0 & 2.0\\
1999dq & 0.51 & 0.9 & -- & 1999ee & 0.59 & 0.9 & --\\
1999gp & 0.81 & 0.9 & -- & 2000E & 0.36 & 1.1 & --\\
2000cx & -- & 1.0 & 8.2 & 2001bt & 0.42 & 1.2 & --\\
2001el & 0.43 & 1.1 & -- & 2002bo & 0.41 & 1.1 & 5.7\\
2002dj & -- & 1.2 & 6.1 & 2002er & 0.38 & 1.2 & 4.1\\
2003cg & -- & 1.3 & 6.9 & 2003du & 0.52 & 1.1 & 5.9\\
2003gs & -- & 1.8 & 3.7 & 2004eo & -- & 1.5 & 4.8\\
2005cf & -- & 1.1 & 5.9 & 2005ke & -- & 1.8 & 2.2\\
2006E & -- & -- & 5.4 & 2006X & -- & 1.1 & 5.4\\
2006ce & -- & -- & 5.9 & 2007af & -- & 1.2 & 5.6\\
2007gi & -- & -- & 7.5 & 2007le & -- & 1.0 & 6.7\\
2007sr & -- & 1.1 & 4.8 & 2008Q & -- & 1.2 & 7.1\\
2011by & -- & 1.1 & 5.5 &  &  &  & \\
\hline
\end{tabular}
\end{center}
\label{tbl:summary_observations}

\noindent{}{Table 2.-- Summary of the analyzed SNe Ia sample. Columns 1,5: SNe Ia name; Column 2,6: estimated $^{56}$Ni mass; Column 3,7: $\Delta m_{15}(B)$ (the decline in $B$-band magnitude during the first $15$ days after peak); Column 4,8: estimated distribution of widths of the ($^{56}$Ni mass-weighted) LOS velocities ($v_{\textrm{mod}}$, defined in the text).}
\end{table}

\end{document}